# Chapter 1 Theory of size, confinement, and oxidation effects

Chang Q. Sun[*]

*School of Electrical and Electronic Engineering, Nanyang Technological University, Singapore 639798 and Institute of Advanced Materials Physics and Faculty of Science, Tianjin University, 300072, P. R. China*

---

[*] E-mail: ecqsun@ntu.edu.sg; URL: www.ntu.edu.sg/home/ecqsun/

I   Introduction

Both oxidation and size reduction form the essential entities that dictate the behaviour of a nanostructured oxide. The atomic and electronic processes of oxidation destroy the initially metallic bond to create new kinds of bonds that determine the behavior of the oxides to vary from their parent metals.[1] Oxygen interaction with atoms of metals relates to the technical processes of corrosion, bulk oxidation, and heterogeneous catalysis, etc.[2] Studies of these processes laid the foundations for applications in microelectronics (gate devices and deep submicron integrated circuit technologies), photo-electronics (photoluminescence, photo-conductance and field emission), magneto-electronics (superconductivity and colossal magneto-resistance) and dielectrics (ferro-, piezo-, pyro-electrics).[3] Involvement of interatomic interaction causes the performance of a solid, or a cluster of atoms, to vary from that of an isolated atom. Adjustment of the relative number of the lower-coordinated surface atoms forms an additional freedom that allows one to tune the properties of a nanosolid with respect to that of its bulk counterpart.  Properties of nanosolids determined by their shapes and sizes are indeed fascinating, which form the basis of the emerging field of nanoscience and nanotechnology that has been recognized as the key significance in science, technology, and economics in the 21st century.

How do the interatomic bonding and the surface atomic portion govern the behavior of a nanosolid? What is the consequence of oxidation? How to make use of the joint effect of size reduction and oxidation in predictable design of oxide nanomaterials?

The size induced property change of nanostructures has inspired tremendous theoretical interest. For instance, a number of models have been developed to explain how the size reduction could induce the blue shift in the photoluminescence (PL) of nanosemiconductors. An *impurity luminescent center* model[4] assumed that the PL blue shift arise from different types of impurity centers in the solid and suggested that the density and types of the impurity centers vary with particle size. S*urface states and surface alloying* mechanism[5,6] proposed that the PL blue shift originate from the extent of surface passivation that is subject to the processing parameters, aging conditions, and operation temperatures.[7] The model of *Inter-cluster interaction and oxidation*[8] also claimed the responsibility for the PL blue shift. The most elegant model for the PL blue shift could be the "quantum confinement (QC)" theory.[9,10,11,12,13] According to the QC theory, the PL energy corresponds to the band gap expansion dictated by electron-hole (e-h) pair (or exciton) production:

$$E_G(R) - E_G(\infty) = \pi^2 \hbar^2 / (2\mu R^2) - 1.786 e^2 / (\varepsilon_r R) + 0.284 E_R \tag{1}$$

where $\mu = m_h^* m_e^* / (m_h^* + m_e^*)$, being the reduced mass of the e-h pair, is an adjustable parameter. The $E_G$ expansion originates from the addition of the kinetic energy, $E_K$, and the Coulomb interaction, $E_p$, of the e-h pairs that are separated by a distance of the particle radius, R, and contribution of the Rydberg or spatial correlation (electron-electron interaction) energy $E_R$ for the bulk semiconductor. The effective dielectric constant $\varepsilon_r$ and the effective mass, $\mu$, describe the effect of the homogeneous medium in the quantum box, which is simplified as a mono-trapping central potential by extending the dimension of a single atom, $d_0$, to that of the solid, D. According to the QC theory, electrons in the conduction band and holes in the valence band are confined spatially by the potential barrier of the surface, or trapped by the potential well of the quantum box. Because of the confinement of both the electrons and the holes, the lowest energy optical transition from the valence to the conduction band increases in energy, effectively increasing the $E_G$. The sum of the kinetic and potential energy of the freely moving carriers is responsible for the $E_G$ expansion, and therefore, the width of the confined $E_G$ grows as the characteristic dimensions of the crystallites decrease.

In contrast, a free-exciton collision model[14] suggested that the $E_G$ expansion arises from the contribution of thermally activated phonons in the grain boundaries rather than the QC effect. During PL measurement, the excitation laser heats the free excitons that then collide with the boundaries of the nanometer-sized fragments. The laser heating the free-excitons up to the temperature in excess of the activation energy required for the self-trapping gives rise to the extremely hot self-trapping excitons (STE's). Because the resulting temperature of the STE's is much higher than the lattice temperature, the cooling of the STE's is dominated by the emission of phonons. However, if the STE temperature comes into equilibrium with the lattice temperature, the absorption of lattice phonons becomes possible. As a result, the blue shift of the STE-PL band is suggested to originate from the activation of hot-phonon-assisted electronic transitions. The blue shift of the STE-PL band depends on the temperature of laser-heated free-excitons that in turn is determined by the size of nano-fragments. This event happens because the temperature (kinetic energy) of the laser-heated free-exciton increases with the number of boundary collisions, which tends to be higher with decreasing size of the nano-fragments. The energy gained from laser heating of the exciton increases with decreasing nanosolid size in an exp(1/R) way.

Another typical issue of nanostructures is their thermal stability. The melting point ($T_m$) of an isolated nanosolid, or a system with weakly linked nanoparticles, drops with solid size (called as supercooling), while the $T_m$ may rise (called as superheating) for an embedded nano-system due to the interfacial effect. The $T_m$ is characterized by the Lindermann's criterion[15] of atomic vibration abruption or Born's criterion[16] of shear modulus disappearance at the $T_m$. The $T_m$ elevation or suppression or the mode of melting in the nanometer regime have been described with the following models: (i) homogeneous melting and growth;[17,18] (ii) random fluctuation melting;[19] (ii) liquid shell nucleation and growth;[20,21,22,23] (iii) liquid-drop[24] formation; (iv) lattice-vibrational instability[25,26] and; (v) surface-phonon instability.[27,28]

The challenge is how to correlate these outstanding mechanisms to the effect of bond order loss of the lower-coordinated surface atoms or the effect of confinement. The origin for one single phenomenon must be intrinsically common to others caused by the same origin of size reduction. Understanding the mechanisms for both size reduction and oxidation and their joint contribution is critical to understanding the behavior of nanostructured oxide materials and related devices.

This chapter describes a bond order-length-strength (BOLS) correlation mechanism[29] for the effect of physical size (bond order loss) and a chemical-bond -- valence-band -- potential-barrier (BBB) correlation mechanism[1] for the effect of oxidation (bond nature alteration) on the performance of a nanostructured oxide. The BBB correlation indicates the essentiality of sp-orbital hybridization of an oxygen atom upon reacting with atoms in solid phase. In the process of oxidation, electronic holes, non-bonding lone pairs, lone-par induced anti-bonding dipoles, and hydrogen bond like are involved through charge transportation and polarization, which dictate the performance of an oxide. Charge transport from metal to oxygen creates the band gap, which turns the metal to be a semiconductor or an insulator; lone pair induced charge polarization lowers the work function of the surface while hydrogen bond like formation due to overdosing with the oxygen additives restores the work function. The often-overlooked events of nonbonding and antibonding are expected to play significant roles in the functioning of an oxide.

The BOLS correlation mechanism indicates the significance of bond order loss of an atom at site surrounding a defect or near the edge of a surface or in an amorphous phase in which the coordination (CN) reduction (deviation of bond order, length, and angle) distributes randomly. Bond order loss causes the remaining bonds of the lower-coordinated atom to contract spontaneously associated with bond-strength gain or atomic potential well depression, which localize electrons and enhance the density of charge, mass, and energy in the relaxed region. The energy density rise in the relaxed region perturbs the Hamiltonian and the associated properties such as the band-gap width, core-level energy, Stokes shift (electron-phonon interaction), and dielectric susceptibility. On the other hand, bond-order loss lowers the cohesive energy of the lower-coordinated atom, which dictates the thermodynamic process such as self-assembly growth, atomic vibration, thermal stability, and activation energies for atomic dislocation, diffusion and chemical reaction.

Consistent insight and numerical agreement with observations evidence the impact of bond nature alteration in oxidation and bond order loss in size reduction and the essentiality and validity of the corresponding BBB and BOLS correlation mechanisms in describing oxide nanomaterials.

## 2  Effects of size and confinement

2.1  Basic concepts

### 2.1.1   Intraatomic trapping

Electrons of a single atom confined by the intra-atomic trapping potential, $V_{atom}(r)$, move around the central ion core in a standing-wave form inside the potential well. The $V_{atom}(r)$ takes a value that varies from several eV to infinity, depending on the orbitals in which electrons are revolving.   The Hamiltonian and the corresponding eigen wave functions and the eigen energies for an electron in the isolated atom are given as:

$$\hat{H}_0 = -\frac{\hbar^2 \nabla^2}{2m} + V_{atom}(r)$$

$$\phi_v(r) \propto \sin^2(k_n r), \text{ and}$$

$$E(n) = \hbar^2 k_k^2 / 2m_e ; k_n = 2n\pi/d_0^2, n = 1,2,3,...$$

the atomic diameter $d_0$ corresponds to the dimension of the potential well of the atom. The branch numbers ($n$) correspond to different energy levels. The energy separation between the nearest two levels depends on $(n+1)^2 - n^2 = 2n+1$.

### 2.1.2   Interatomic bonding and intercluster coupling

When two atoms or more joined as a whole, interatomic interaction comes into play, which causes the performance of a cluster of atoms from that of an isolated atom. The interatomic bonding is essential to make a solid or even a liquid. Considering an assembly composed of $n$ particles of mean size $K_j$ and with each particle, there are $N_j$ atoms, the total binding energy, $V_{cry}(r, n, N_j)$, is:[30]

$$\begin{aligned} V_{cry}(r,n,N_j) &= \sum_n \sum_{l \neq i} \sum_i v(r_{li}) \\ &= \frac{n}{2}\left[N_j \sum_{i=1} v(r_{li}) + \sum_{k \neq j} V(K_{kj})\right] \\ &\cong \frac{n}{2}\left[N_j^2 v(d_0) + nV(K_j)\right] \end{aligned}$$

(2)

The $V_{cry}(r, n, N_j)$ sums over all the $N_j$ atoms and the n particles. The high order $r_{li}$ is a certain fold of the nearest atomic spacing, $d_0$. Interaction between the nearest clusters, $k$ and $j$, $V(K_{kj})$, is negligible if the $K_{kj}$ is considerably large. Normally, the intercluster interaction, $V(K_{kj})$, is much weaker than the interatomic interaction, if the cluster is treated as an electric or a magnetic dipole of which the Van der Waals or the super-paramagnetic potentials dominate.

### 2.1.3 Hamiltonian and energy band

According to the band theory, the Hamiltonian for an electron inside a solid is in the form:

$$\hat{H} = \hat{H}_0 + \hat{H}' = -\frac{\hbar^2 \nabla^2}{2m} + V_{atom}(r) + V_{cry}(r + R_C)$$

(3)

where the $\hat{H}' = V_{cry}(r) = V_{cry}(r + R_C)$ is the periodic potential of the crystal. $R_C$ is the lattice constant. According to the nearly-free-electron approximation, the $E_G$ between the valence and the conduction bads originates from the crystal potential. The width of the gap depends on the integral of the crystal potential in combination with the Bloch wave of the nearly free electron, $\phi(k_l, r)$:

$$E_G = 2|V_1(k_l)|, \text{ and } V_1(k_l) = \langle \phi(k_l, r) | V(r + R_C) | \phi(k_l, r) \rangle$$

(4)

where $k_l$ is the wave-vector and $k_l = 2l\pi/R_C$. Actually, the $E_G$ is simply twice the first Fourier coefficient of the crystal potential.

As illustrated in Figure 1, the energy levels of an isolated atom will evolve into energy bands when interatomic bonding is involved. When a system contains two atoms, such as a dimer like $H_2$ molecules, the single energy level splits into two sublevels. The presence of interatomic interaction lowers the center of the two sublevels, which is termed as core level shift. Increasing the number of atoms up to $N_j$, the single energy level will expand into a band within which there are $N_j$ sublevels.

What distinguishes a nanosolid from a bulk chunk is that for the former the $N_j$ is accountable, while for the latter the $N_j$ is too large to be accounted despite the portion of the lower-coordinated atoms in the surface skin. Therefore, the classical band theories are valid for a solid that may contain any number of atoms. As detected with x-ray photoelectron spectroscopy (XPS), the density-of-states (DOS) of a core band for a nanosolid exhibits band-like features rather than the discrete spectral lines of a single atom. If the $N_j$ is sufficiently small, the separation between the sublevels is resolvable. The energy level spacing between the successive sublevels in the valence band, know as the Kubo gap ($\delta_K$ = $4E_F/3N_j$), decreases with increasing the number of valence electrons of the system, $N_j$.[31] Where $E_F$ is the Fermi energy of the bulk. Because of the presence of the $\delta_K$ in an individual nanosolid, properties such as electron conductivity and magnetic susceptibility exhibit quantized features.[32]

Figure 1 The involvement of interatomic interaction evolves a single energy level to the energy band when a particle grows from a single atom to a bulk solid that contains $N_j$ atoms. Indicated are the work function $\phi$, band gap $E_G$, core level shift $\Delta E_\nu$, bandwidth $E_B$. The number of allowed sublevels in a certain band equals the number of atoms of the solid.[33]

According to the tight-binding approximation, the energy dispersion of an electron in the ν th core band follows the relation:

$$\begin{aligned} E_\nu(k) &= E_\nu(1) + \Delta E_\nu(\infty) + \Delta E_B(k_l, R_C, z) \\ &= E_\nu(1) - (\beta + 2\alpha) + 4\alpha\Omega(k_l, R_C, z) \end{aligned}$$

(5)

where,

$E_\nu(1) = \langle \phi_\nu(r) | \hat{H}_0 | \phi_\nu(r) \rangle$ is the ν th energy level of an isolated atom.

$\beta = -\langle \phi_\nu(r) | V_{cry}(r) | \phi_\nu(r) \rangle$ is the crystal potential effect on the specific core electron at site r.

$\alpha = -\langle \phi_\nu(r - R_C) | V_{cry}(r - R_C) | \phi_\nu(r - R_C) \rangle$ is the crystal potential effect on the coordinate neighbouring electrons. For an fcc structure example, the structure factor, $\Omega(k_l, R_C) = \sum_z \sin^2(k_l R_C / 2)$. The sum is over all the contributing coordinates (z) surrounding the specific atom in the solid.

Eqs (4) and (5) indicate that the $E_G$, the energy shift $\Delta E_\nu(\infty) = -(\beta + 2\alpha)$ of the $E_\nu(1)$ and the bandwidth $\Delta E_B$ (last term in Eq (5)) are all functions of the crystal potential. Any perturbation to the crystal potential will vary these quantities accordingly as the change of the Block wavefunction is negligible in the first order approximation. The band structure has nothing to do with the actual occupancy of the particular orbitals or events such as electron-hole pair creation or recombination, or the processes of PL and PA that involve electron-phonon coupling effect. Without the crystal potential, neither the $E_G$ expansion nor the core-level shift would be possible; without the inter-atomic binding, neither a solid nor even a liquid would form.

If one intends to modify the properties of a solid, one has to find ways of modulating the crystal potential physically or chemically. Bond nature alteration by chemical reaction or bond length relaxation by size reduction, as discussed in the following sections, will be the effective ways of modulating the interatomic potential.

### 2.1.4 Atomic cohesive energy and thermal stability

Another key concept is the cohesive energy per discrete atom. The binding energy density per unit volume contributes to the Hamiltonian that determines the entire band structure and related properties, while the atomic cohesive energy determines the activation energy for thermally and mechanically activated processes including self assembly growth, phase transition, solid-liquid transition, evaporation, atomic dislocation, diffusion and chemical reaction.

The cohesive energy ($E_{coh}$) of a solid containing $N_j$ atoms equals to the energy dividing the crystal into individually isolated atoms by breaking all the bonds of the solid. If no atomic CN reduction is considered, the $E_{coh}$ is the sum of bond energy over all the $z_b$ coordinates of all the $N_j$ atoms:

$$E_{coh}(N_j) = \sum_{N_j} \sum_{z_i} E_i \cong N_j z_b E_b = N_j E_B$$

**(6)**

The cohesive energy for a single atom, $E_B$, is the sum of the single bond energy $E_b$ over the atomic *CN*, $E_B = z_b E_b$. One may consider a thermally activated process such as phase transition in which all the bonds are loosened to a certain extent due to thermal activation. The energy required for such a process is a certain portion of the atomic $E_B$ though the exact portion may change from process to process. If one considers the relative change to the bulk value, the portion will not be accounted. This approximation is convienent in practice, as one should be concerned with the origins and the trends of changes. Therefore, bulk properties such as thermal stability of a solid could be related directly to the atomic cohesive energy – the product of the bond number and bond energy of the specific atom.

2.2 Boundary conditions

2.2.1  Barrier confinement vs. quntum uncertainty

The termination of lattice periodicity in the surface normal direction has two effects. One is the creation of the surface potential barrier (SPB), or work function, or contact potential, and the other is the reduction of the atomic CN. The SPB is the intrinsic feature of a surface, which confines only electrons that are freely moving inside the solid. However, the SPB has nothing to do with the *strongly localized* electrons in deeper core bands or with those form sharing electron pairs in a bond. The localized electrons do not suffer such barrier confinement at all as the localization length is far shorter than the particle size.

According to the quantum uncertainty principle, reducing the dimension (D) of the space inside which energetic particles are moving increases the fluctuation, rather than the average value, of the momentum, p, or kinetic energy, $E_k$, of the moving particles:

$$\Delta p D \geq \hbar/2$$
$$p = \bar{p} \pm \Delta p$$
$$\overline{E_k} = \bar{p}^2/(2\mu)$$

(7)

where $\hbar$ being the Plank constant corresponds to the minimal quanta in energy and momentum spaces and $\mu$ is the mass of the moving particle. The kinetic energy of a freely moving particle is increased by a negligible amount due to the confinement effect that follows the quantum uncertainty principle. If the average momentum $\bar{p} = 0$, then $\Delta p \cong p$, the kinetic energy enhancement would be: $\overline{E_k} = \Delta p^2/(2\mu) = (\hbar/2D)^2/2\mu \approx D^{-2}10^{-18}(eV)$. If the D is taken as 10 nm, the $E_k$ is increased by $10^{-2}$ eV. Therefore, the SPB confinement causes energy-rise of neither the freely moving carriers nor the localized ones.

### 2.2.2 Atomic CN reduction

The atomic CN reduction is referred to the standard value of 12 in the bulk of an fcc structure irrespective of the bond nature or the crystal structure. Atomic CN reduction is refereed to an atom with coordinate less than the standard value of 12. The CN is 2 for an atom in the interior of a monatomic chain or an atom at the open end of a single-walled carbon nanotube (CNT); while in the CNT wall, the CN is 3. For an atom in the fcc unit cell, the CN varies from site to site. The CN of an atom at the edge or corner differs from the CN of an atom in the plane or the central of the unit cell. Atoms with deformed bond lengths or deviated angles in the CNT are the same as those in amorphous states that are characterised with the band tail states.[34] For example, the effective CN of an atom in diamond tetrahedron is the same as that in an fcc structure as a tetrahedron unit cell is an interlock of two fcc unit cells. The CN of an atom in a highly curved surface is even lower compared with the CN of an atom at a flat surface. For a negatively curved surface (such as the inner side of a pore or a bubble), the CN may be slightly higher than that of an atom at the flat surface. Therefore, from the atomic CN reduction point of view, there is no substantial difference in nature between a nanosolid, a nanopore, and a flat surface. This premise can be extended to the structural defects or defaults such as voids surrounding which atoms are suffer from CN reduction. Unlike a nansolid with ordered CN reduction at the surface, an amorphous solid possesses defects that are distributed randomly.

### 2.3 Surface-to-volume ratio

It is easy to derive the volume or number ratio of a certain atomic layer, denoted i, to that of the entire solid by differentiating the natural logrithum of the volume,

$$\gamma_{ij} = \frac{N_i}{N_j} = \frac{V_i}{V_j} = dLn(V_j) = \frac{\tau dR_j}{R_j} = \frac{\tau c_i}{K_j}$$

(8)

where $K_j = R_j/d_0$ is the dimensionless form of size, which is the number of atoms lined along the radius of a spherical dot ($\tau = 3$), a rod ($\tau = 2$), or cross the thickness of a thin plate ($\tau = 1$). The volume of a solid is proportional to $R_j^\tau$. For a hollow system, the $\gamma_{ij}$ should count both external and internal sides of the hollow structure.

With reducing particle size, performance of surface atoms become dominant because at the lower end of size limit ($K_j \rightarrow \tau c_i$) $\gamma_1$ approaches unity. At $K_j = 1$, the solid will degenerate into an isolated atom. Therefore, the $\gamma_{ij}$ covers the whole range of sizes and various shapes. The definition of dimensionality ($\tau$) herein differs from the convention in transport or quantum confinement considerations in which a nano-sphere is zero-dimension (quantum dot), a rod as one dimension (quantum wire), and a plate two dimension (quantum well). If we count atom by atom, the number ratio and the property change will show quantized oscillation features at smaller sizes, which varies from structure to structure.[35]

2.4 Bond order-length-strength (BOLS) correlation

2.4.1 Bond order-length correlation

As the consequence of bond order loss, the remaining bonds of the lower-coordinated atoms contract spontaneously. As aserted by Goldschmidt[36] and Pauling,[37] the ionic and the metallic radius of the atom would shrink spontaneously if the CN of an atom is reduced. The CN reduction induced bond contraction is independent of the nature of the specific bond or structural phases.[38] A 10% contraction of spacing between the first and second atomic surface layers has been detected in the liquid phase of Sn, Hg, Ga, and In.[39] As impurity has induced 8% bond contraction around the impurity (acceptor dopant As) at the Te sublattice in CdTe has also been observed using EXAFS (extented X-ray absorption fine structure) and XANES (X-ray absorption near edge spectroscopy).[40] Therefore, CN reduction induced bond contraction is common.

Figure 2a illustrates the CN dependence of bond length. The solid curve, $c_i(z_i)$, formulates the Goldschmidt premise which states that an ionic radius contracts by 12%, 4%, and 3%, if the CN of the atom reduces from 12 to 4, 6 and 8, respectively. Feibelman[41] has noted a 30% contraction of the dimer bond of Ti and Zr, and a 40% contraction of the dimer-bond of Vanadium, which is also in line with the formulation.

Figure 2 (link) illustration of the BOLS correlation. Solid curve in (a) is the contraction coefficient $c_i$ derived from the notations of Goldschmidt[36] (open circles) and Feibelman[41] (open square). As a spontaneous process of bond contraction, the bond energy at equilibrium atomic separation will rise in absolute energy, $E_i = c_i^{-m} E_b$. The $m$ is a parameter that represents the nature of the bond. However, the atomic cohesive energy, $z_i E_i$, changes with both the m and $z_i$ values. (b) Atomic CN reduction modified pairing potential energy. CN reduction causes the bond to contract from one unit (in $d_0$) to $c_i$ and the cohesive energy per coordinate increases from one unit to $c_i^{-m}$ unit. Separation between $E_i(T)$ and $E_i(0)$ is the thermal vibration energy. Separation between $E_i(T_{m,i})$ and $E_i(T)$ corresponds to melting energy per bond at T, which dominates the mechanical strength. $T_{m,i}$ is the melting point. $\eta_{2i}$ is $1/z_i$ fold energy atomizing an atom in molten state.

2.4.2 Bond length-strength correlation

As a consequence of the spontaneous process of bond contraction, the bond strength will gain, towards a lowering of the system energy. The contraction coefficient and the associated bond energy gain form the subject of the BOLS correlation mechanism that is formulated as:

$$\begin{cases} c_i(z_i) = d_i/d_0 = 2/\{1+\exp[(12-z_i)/(8z_i)]\} & (BOLS-coefficient) \\ E_i = c_i^{-m} E_b & (Single-bond-energy) \\ E_{B,i} = z_i E_i & (atomic-cohesive-energy) \end{cases}$$

(9)

Subscript i denotes an atom in the i th atomic layer, which is countered up to three from the outermost atomic layer to the center of the solid as no CN-reduction is expected for i > 3. The index m is a key indicator for the nature of the bond. Experience[42] revealed that for Au, Ag, Ni metals, m ≡ 1; for alloys and compounds, m is around four; for C and Si, the $m$ has been optimized to be 2.56[43] and 4.88,[44] respectively. The m value may vary if the bond nature evolves with atomic CN.[45] If the surface bond expands in cases, we simply expand the $c_i$ from a value that is smaller than unity to greater, and the m value from positive to negative to represent the spontaneous process of which the system energy is minimized. The $c_i(z_i)$ depends on the effective CN rather than a certain order of CN. The $z_i$ also varies with the particle size due to the change of the surface curvature. The $z_i$ takes the following values:[44]

$$z_1 = \begin{cases} 4(1-0.75/K_j) & curved-surface \\ 4 & flat-surface \end{cases}$$

(10)

Generally, $z_2 = 6$ and $z_3 = 8$ or 12.

Figure 2b illustrates schematically the BOLS correlation using a simple interatomic pairing potential, u(r). When the CN of an atom is reduced, the equilibrium atomic distance will contract from one unit (in $d_0$) to $c_i$ and the cohesive energy of the shortened bond will increase in magnitude from one unit (in $E_b$) to $c_i^{-m}$. The solid and the broken u(r) curves correspond to the pairing potential with and without CN reduction. The BOLS correlation has nothing to do with the particular form of the pairing potential as the approach involves only atomic distance at equilibrium. The bond length-strength correlation herein is consistent with the trend reported by Bahn and Jacobsen[46] though the extents of bond contraction and energy enhancement therein vary from situation to situation.

There are several characteristic energies in Figure 2b, which correspond to the following facts:
(i) $T_{m,i}$ being the local melting point is proportional to the cohesive energy, $z_iE_i(0)$,[47] per atom with $z_i$ coordinate.[48]
(ii) Separation between E = 0 and $E_i(T)$, or $\eta_{1i}(T_{m,i} - T) + \eta_{2i}$, corresponds to the cohesive energy per coordinate, $E_i$, at T, being energy required for bond fracture under mechanical or thermal stimulus. $\eta_{1i}$ is the specific heat per coordinate.
(iii) The separation between E = 0 and $E_i(T_m)$, or $\eta_{2i}$, is the $1/z_i$ fold energy that is required for atomization of an atom in molten state.
(iv) The spacing between $E_i(T)$ and $E_i(0)$ is the vibration energy purely due to thermal excitation.
(v) The energy contributing to mechanical strength is the separation between the $E_i(T_m)$ and the $E_i(T)$, as a molten phase is extremely soft and highly compressible.[49]

Values of $\eta_{1i}$ and $\eta_{2i}$ can be obtained with the known $c_i^{-m}$ and the bulk $\eta_{1b}$ and $\eta_{2b}$ values that vary only with crystal structures as given in Table 1.

Table 1 Relation between the bond energy $E_b$ and the $T_m$ of various structures.[24] $\eta_{2b} < 0$ for an fcc structure means that the energy required for breaking all the bonds of an atom in molten state is included in the term of $\eta_{1b}zT_m$ and therefore the $\eta_{2b}$ exaggerates the specific heat per CN.

| $E_b = \eta_{1b}T_m + \eta_{2b}$ | fcc | bcc | Diamond structure |
|---|---|---|---|
| $\eta_{1b}$ ($10^{-4}$ eV/K) | 5.542 | 5.919 | 5.736 |
| $\eta_{2b}$ (eV) | -0.24 | 0.0364 | 1.29 |

### 2.4.3 Densification of mass, charge, and energy

Figure 3 compares the potential well in the QC convention with that of the BOLS for a nanosolid. The QC convention extends the monotrapping potential of an isolated atom by expanding the size from $d_0$ to D. BOLS scheme covers contribution from individual atoms that are described with multi-trapping-centre potential wells, and the effect of atomic CN reduction in the surface skin. Atomic CN reduction induced bond-strength gain depresses the potential well of trapping in the surface skin. Therefore, the density of charge, energy, and mass in the relaxed surface region are higher than other sites inside the solid. Consequently, surface stress that is in the dimension of energy density will increase in the relaxed region. Electrons in the relaxed region are more localized because of the depression of the potential well of trapping, which lowers the work function and conductivity in the surface region, but enhances the angular momentum of the surface atoms.[35]

Figure 3 ([link](#)) Schematic illustration of conventional quantum well (upper part) with a monotrapping center extended from that of a single atom, and the BOLS derived nanosolid potential (lower part) with multi-trap centers and CN reduction induced features. In the relaxed surface region, the density of charge, energy and mass will be higher than other sites due to atomic CN reduction.

## 2.5 Shape-and-size dependency
### 2.5.1 Scaling relation

Generally, the mean relative change of a measurable quantity of a nanosolid containing $N_j$ atoms, with dimension $K_j$, can be expressed as $Q(K_j)$; and as $Q(\infty)$ for the same solid without contribution from bond order loss. The correlation between the $Q(K_j)$ and $Q(\infty) = N_j q_0$ and the relative change of Q due to bond order loss is given as:

$$\begin{aligned} Q(K_j) &= N_j q_0 + N_s(q_s - q_0) \\ \frac{\Delta Q(K_j)}{Q(\infty)} &= \frac{Q(K_j) - Q(\infty)}{Q(\infty)} = \frac{N_s}{N_j}\left(\frac{q_s}{q_0} - 1\right) \\ &= \sum_{i \leq 3} \gamma_{ij}(\Delta q_i / q_0) = \Delta_{qj} \end{aligned}$$

(11)

The weighting factor, $\gamma_{ij}$, represents the geometrical contributions from dimension ($K_j$) and dimensionality ($\tau$) of the solid, which determines the magnitude of change. The quantity $\Delta q_i/q_0$ is the origin of change. The $\sum_{i \leq 3} \gamma_{ij}$ drops in a $K_j^{-1}$ fashion from unity to infinitely small when the solid grows from atomic level to infinitely large. For a spherical dot at the lower end of the size limit, $K_j = 1.5$ ($K_j d_0 = 0.43$ nm for an Au spherical dot example), $z_1 = 2$, $\gamma_{1j} = 1$, and $\gamma_{2j} = \gamma_{3j} = 0$, which is identical in situation to an atom in a monatomic chain (MC) despite the orientation of the two interatomic bonds. Actually, the bond orientation is not involved in the modeling iteration. Therefore, the performance of an atom in the smallest nanosolid is a mimic of an atom in an MC of the same element without presence of external stimulus such as stretching or heating. At the lower end of the size limit, the property change of a nanosolid relates directly to the behavior of a single bond.

Generally, experimentally observed size-and-shape dependence of a detectable quantity follows a scaling relation. Equilibrating the scaling relation to Eq (**9**), one has:

$$Q(K_j) - Q(\infty) = \begin{cases} b K_j^{-1} & (measurement) \\ Q(\infty) \times \Delta_{qj} & (theory) \end{cases}$$

(12)

where the slope $b \equiv Q(\infty) \times \Delta_{qj} \times K_j \cong$ constant is the focus of various modeling pursues. The $\Delta_j \propto K_j^{-1}$ varies simply with the $\gamma_{ij}(\tau, K_j, c_i)$ if the functional dependence of $q(z_i, c_i, m)$ on the atomic CN, bond length, and bond energy is given.

### 2.5.2  Cohesive energy modification

The heat energy required for loosening an atom is a certain portion of the atomic $E_B$ that varies with not only the atomic CN but also the bond strength. The variation of the mean $E_B$ with size is responsible for the fall (supercooling) or rise (superheating) of the $T_C$ (critical temperature for melting, phase transition, or evaporation) of a surface and a nanosolid. The $E_B$ is also responsible for other thermally activated behaviors such as phase transition, catalytic reactivity, crystal structural stability, alloy formation (segregation and diffusion), and stability of electrically charged particles (Coulomb explosion). The cohesive energy also determines crystal growth and atomic diffusion, and atomic gliding displacement that determine the ductility of nanosolids.

The BOLS correlation considers contribution from atoms in the shells of the surface skin. Using the spherical dot containing $N_j$ atoms, the average $<E_{coh}(N_j)>$, or $<E_B(N_j)>$ is,

$$\langle E_{coh}(N_j) \rangle = N_j z_b E_b + \sum_{i \leq 3} N_i (z_i E_i - z_b E_b)$$

$$= N_j E_B(\infty) + \sum_{i \leq 3} N_i z_b E_b (z_{ib} E_{ib} - 1)$$

$$= E_{coh}(\infty) \left[ 1 + \sum_{i \leq 3} \gamma_{ij} (z_{ib} c_i^{-m} - 1) \right] = E_{coh}(\infty)(1 + \Delta_B)$$

$$\text{or,} \quad \Delta E_B(K_j)/E_B(\infty) = \sum_{i \leq 3} \gamma_{ij} (z_{ib} c_i^{-m} - 1) = \Delta_B$$

(13)

where $E_{coh}(\infty) = N_j z_b E_b$ represents the ideal situation without *CN reduction*. The $z_{ib} = z_i/z_b$ is the normalized *CN* and $E_{ib} = E_i/E_b \cong c_i^{-m}$ is the normalized binding energy per coordinate of a surface atom. For an isolated surface, $\Delta_B < 0$; for an intermixed interface, $\Delta_B$ may be positive depending on the strength of interfacial interaction. Therefore, the relative change of $T_C(K_j)$ and activation energy, $E_A(K_j)$ for thermally and mechanically activated process can be expressed as:

$$\frac{\Delta T_C(K_j)}{T_C(\infty)} = \frac{\Delta E_A(K_j)}{E_A(\infty)} = \frac{\Delta E_B(K_j)}{E_B(\infty)} = \Delta_B(K_j)$$

(14)

Interestingly, the critical temperature for sensoring operation could be lowered from 970 to 310 K of SrTiO$_3$ by ball milling to obtain 27 nm-sized powders.[50] The resistivity of the SrTiO$_3$ increases when the SrTiO$_3$ particle size is decreased.[51] Decreasing the particle sizes of ferroelectric BaTiO$_3$ could lower the T$_C$ to 400 K and the refractive index (dielectric constant), and hence the transmittance of BaTiO$_3$ infilled SiO$_2$ photonic crystals, as a consequence.[52,53] The suppression of the critical temperatures for sensoring and phase transition results and the modulation of resistivity and refractive index could be consequences of energy densification and cohesive energy suppression in the surface skin.

2.5.3    Hamiltonian perturbation

The perturbation to the energy density in the relaxed region that contributes to the Hamiltonian upon assembly of the nanosolids is,

$$\Delta_H(K_j) = \frac{V_{cry}(r, n, N_j)}{V_{cry}(d_0, n, N_j)} - 1$$

$$= \sum_{i \leq 3} \gamma_{ij} \frac{\Delta v(d_i)}{v(d_0)} + \delta_{kj}$$

$$= \sum_{i \leq 3} \gamma_{ij} (c_i^{-m} - 1) + \delta_{kj}$$

$$\text{where, } \delta_{kj} = \frac{nV(K_j)}{N_j^2 v(d_0)}$$

(15)

With the perturbation, the $\hat{H}'$ in eq (3) becomes, $\hat{H}'(\Delta_H) = V_{cry}(r)[1 + \Delta_H(K_j)]$, which dictates the change of not only the $E_G$ width, but also the core-level energy:

$$\frac{\Delta E_G(K_j)}{E_G(\infty)} = \frac{\Delta E_\nu(K_j)}{E_\nu(\infty)} = \Delta_H(K_j)$$

**(16)**

where $\Delta E_\nu(K_j) = E_\nu(K_j) - E_\nu(1)$. This relation also applies to other quantities such as the bandwidth and band tails.[33]

Most strikingly, without triggering electron-phonon interaction or electron-hole generation, scanning tunneling microscopy/spectroscopy (STM/S) measurement at low temperature revealed that the $E_G$ of Si nanorod varies from 1.1 to 3.5 eV with decreasing the rod diameter from 7.0 to 1.3 nm associated with ~12% Si-Si bond contraction from the bulk value (0.263 nm) to ~0.23 nm. The STS findings concur excitingly with the BOLS premise: CN reduction shortens the remaining bonds of the lower-coordinated atoms spontaneously with an association of $E_G$ expansion.

### 2.5.4 Electron-phonon coupling

Electron-phonon (e-p) interaction contributes to the processes of photoemission, photoabsorption, photoconduction, and electron polarization that dominates the static dielectric constant. Figure 4 illustrates the effect of e-p coupling and crystal binding on the energy of photoluminescence and adsorbace, $E_{PL}$ and $E_{PA}$. The energies of the ground state ($E_1$) and the excited state ($E_2$) are expressed in parabola forms:[34]

$$\begin{cases} E_1(q) = Aq^2 \\ E_2(q) = A(q - q_0)^2 + E_G \end{cases}$$

(17)

Constant A is the slope of the parabolas. The $q$ is in the dimension of wave-vector. The vertical distance between the two minima is the true $E_G$ that depends uniquely on the crystal potential. The lateral displacement ($q_0$) originates from the e-p coupling that can be strengthened by lattice contraction. Therefore, the blue shift in the $E_{PL}$ and in the $E_{PA}$ is the joint contribution from crystal binding and e-p coupling.

Figure 4 (link) Mechanisms for $E_{PA}$ and $E_{PL}$ of a nano-semiconductor, involving crystal binding ($E_G$) and electron-phonon coupling (W). Insertion illustrates the Stokes shift from $E_{PA}$ to $E_{PL}$. Electron is excited by absorbing a photon with energy $E_G+W$ from the ground minimum to the excited state and then undergoes a thermalization to the excited minimum, and then transmits to the ground emitting a photon with energy $E_G-W$.[54]

In the process of carrier formation, or electron polarization,[34] an electron is excited by absorbing a photon with $E_G+W$ energy from the ground minimum to the excited state with creation of an electron-hole pair. The excited electron then undergoes a thermalization and moves to the minimum of the excited state, and eventually transmits to the ground and combines with the hole. The carrier recombination is associated with emission of a photon with energy $E_{PL} = E_G - W$. The transition processes (e-h pair production and recombination) follow the rule of momentum and energy conservation though the conservation law may be subject to relaxation for the short ordered nanosolid. Relaxation of the conservation law is responsible for the broad peaks in the PA and PL.

The insertion illustrates the Stokes shift, $2W = 2Aq_0^2$, or the separation from $E_{PL}$ to $E_{PA}$. The $q_0$ is inversely proportional to atomic distance $d_i$, and hence, $W_i = A/(c_i d_i)^2$, in the surface region. Based on this premise, the blue shift of the $E_{PL}$, the $E_{PA}$, and the Stokes shift can be correlated to the CN reduction-induced bond contraction:[54]

$$\left.\begin{array}{l}\dfrac{\Delta E_{PL}(K_j)}{E_{PL}(\infty)} \\ \dfrac{\Delta E_{PA}(K_j)}{E_{PA}(\infty)}\end{array}\right\} = \dfrac{\Delta E_G(K_j) \mp \Delta W(K_j)}{E_G(\infty) \mp W(\infty)} \cong \sum_{i \leq 3} \gamma_i \left[ (c_i^{-m} - 1) \mp B(c_i^{-2} - 1) \right]$$

$$= \Delta_H \mp B\Delta_{e-p}$$

$$\left( B = \dfrac{A}{E_G(\infty) d^2}; \quad \dfrac{W(\infty)}{E_G(\infty)} \approx \dfrac{0.007}{1.12} \approx 0 \right)$$

(18)

Compared with the bulk $E_G(\infty) = 1.12$ eV for silicon, the $W(\infty) \sim 0.007$ eV obtained using tight-binding calculations[55] is negligible. One can easily calculate the size dependent $E_{PL}$, $E_{PA}$, and $E_G = (E_{PL} + E_{PA})/2$ using Eq (18). Fitting the measured data gives the values of *m* and *A* for a specific semiconductor.

2.5.5    Mechanical strength

The mechanical yield strength is the strain-induced internal energy deviation that is proportional to energy density or the sum of bond energy per unit volume.[1] Considering the contribution from heating, the strength (stress, flow strength), the Young's modulus, and the compressibility (under compressive stress) or extensibility (under tensile stress) at a given temperature can be expressed by:

$$P_i(z_i, T) = -\left.\frac{\partial u(r,T)}{\partial V}\right|_{d_i,T} \sim \frac{N_i \eta_{1i}(T_{m,i} - T)}{d_i^\tau}$$

$$\beta_i(z_i, T) = -\left.\frac{\partial V}{V \partial P}\right|_T = [Y_i(z_i, T)]^{-1} = \left[-V \left.\frac{\partial u^{2^{-1}}(r,T)}{\partial V^2}\right|_T\right]^{-1}$$

$$= \frac{d_i^\tau}{N_i \eta_{1i}(T_{m,i} - T)} = [P_i(z_i, T)]^{-1}$$

(19)

β is an inverse of dimension of the Young's modulus or the hardness. $N_j$ is the total number of bonds in $d^\tau$ volume. If calibrated with the bulk value at T and using the size dependent specific heat, melting point, and lattice parameter, the temperature, bond nature, and size dependent strength and compressibility of a nanosolid will be:

$$\frac{P(K_j, T)}{P(\infty, T)} = \frac{Y(K_j, T)}{Y(\infty, T)} = \frac{\beta(\infty, T)}{\beta(K_j, T)} = \frac{\eta_1(K_j)}{\eta_1(\infty)}\left(\frac{d(\infty)}{d(K_j)}\right)^3 \times \frac{T_m(K_j, m) - T}{T_m(\infty) - T}$$

(20)

The bond number density between the circumferential neighbouring atomic layers does not change upon relaxation ($N_i = N_b$). Eq (**20**) indicates that the mechanical strength is dictated by the value of $T_m(K_j) - T$ and the specific heat per bond. At T far below the $T_m$, a surface or a nanostructure is harder than the bulk interior. However, the $T_m$ drops with size $K_j$ and therefore, the surface or nanosolid become softer when the $T_m(K_j) - T$ value becomes smaller. This relation has led to quantification of the surface mechanical strength, the breaking limit of a single bond in monatomic chain and the anomalous Hall-Petch relationship in which the mechanical strength decreases with size in a fashion of $D^{-0.5}$ and then deviates at 10 nm from the Hall-Petch relationship.[49]

2.6 Summary

If one could establish the functional dependence of a detectable quantity *Q* on atomic separation or its derivatives, the size dependence of the quantity *Q* is then certain. One can hence design a nanomaterial with desired functions based on such prediction. Physical quantities of a solid can be normally categorized as follows:

(i) Quantities that are directly related to bond length, such as the mean lattice constant, atomic density, and binding energy.

(ii) Quantities that depend on the cohesive energy per discrete atom, $E_{B,i} = z_i E_i$, such as self-organization growth, thermal stability, Coulomb blockade, critical temperature for liquidation, evaporation and phase transition of a nanosolid and the activation energy for atomic dislocation, diffusion, and bond unfolding.[56]

(iii) Properties that vary with the binding energy density in the relaxed continuum region such as the Hamiltonian that determines the entire band structure and related properties such as band-gap, core-level energy, photoabsorption and photoemission.

(iv) Properties that are contributed from the joint effect of the binding energy density and atomic cohesive energy such as mechanical strength, Young's modulus, surface energy, surface stress, extensibility and compressibility of a nanosolid, as well as the magnetic performance of a ferromagnetic nanosolid.

Using the scaling relation and the BOLS correlation, we may derive solutions to predict the size and shape dependence of various properties. Typical samples are given in Table 2 and Figure 5.

Table 2 Summary of functional dependence of various quantities on particle size and derived information. Typical samples of consistency are shown in Figure 5.

| Quantity Q | $\Delta Q(K_j)/Q(\infty) = \Delta_q(K_j)$ | Refs | Comments |
|---|---|---|---|
| Lattice constant (d) | $\sum_{i \leq 3} \gamma_{ij}(c_i - 1)$ | 42 | Only outermost three atomic layers contribute |
| Bond energy ($E_i$) Band gap ($E_G$) Core-level shift ($\Delta E_v$) | $\sum_{i \leq 3} \gamma_{ij}(c_i^{-m} - 1) = \Delta_H$ | 29,33 | $\Delta_H$ - Hamiltonian perturbation |
| Electron-phonon coupling energy (Stokes shift, W) | $B \sum_{i \leq 3} \gamma_{ij}(c_i^{-2} - 1) = B\Delta_{e-p}$ | 54 | B-constant |
| Photoemission and photoabsorption energy ($E_{PL}$, $E_{PA}$) | $\Delta_H \mp B\Delta_{e-p}$ | 54 | $E_G = (E_{PA} + E_{PL})/2$ |
| Critical temperature for phase transition ($T_C$); activation energy for | $\sum_{i \leq 3} \gamma_{ij}(z_{ib} c_i^{-m} - 1) = \Delta_B$ | 47,57 | $\Delta_B$ - Atomic cohesive perturbation |
| thermally and mechanically activated processes | | | |
| Mechanical strength and Yung's modulus of monatomic bond (P, Y) | $\dfrac{\eta_{1i} d^3 (T_{m,i} - T)}{\eta_{1b} d_i^3 (T_m(\infty) - T)} - 1$ | 58 | $\eta_1$ – specific heat per bond $T_m$ – melting point |
| Inverse Hall-Petch relation; solid-semisolid-liquid transition | $\dfrac{\eta_1(K_j) d^3 (T_m(K_j) - T)}{\eta_{1b} d^3(K_j)(T_m(\infty) - T)} - 1 = \begin{cases} \leq 0, & (semisolid) \\ -1, & (liquid) \end{cases}$ | 49 | |
| Optical phonon frequency (ω) | $\sum_{i \leq 3} \gamma_{ij} \left( z_{ib} c_i^{-(m/2+1)} - 1 \right)$ | 59 | |
| Fermi level | $\sum_{i \leq 3} \gamma_{ij} \left( c_i^{-2\tau/3} - 1 \right)$ | 60,61 | |
| Dielectric permittivity ($\chi = \varepsilon_r - 1$) | $\Delta_d - (\Delta_H - B\Delta_{e-p})$ | 62 | |

Figure 5  Comparison of BOLS predictions with measured size dependence of
  (a) Lattice contraction of $Pr_2O_3$ films on Si substrate.[63]
  (b) Atomic cohesive energy of Mo and W. [64]
  (c) Mechanical strength (Inverse hall-Patch relationship, IHPR) of $TiO_2$[65] nanosolids; Straight line is the traditional Hall-Petch relationship (HPR).
  (d) $T_m$ suppression of Bi[66, 67,68,69,70] and CdS,[71] and
  (e) $T_C$ suppression of ferromagnetic $Fe_3O_4$ nanosolids; [72]
  (f) $T_C$ suppression of ferroelectric $PbTiO_3$,[73] $SrBi_2Ta_2O_9$,[74] $BaTiO_3$,[75] and anti-ferroelectric $PbZrO_3$[76] nanosolids. High order CN reduction is considered for dipole-dipole interaction.[57]

Figure 6  Comparison of BOLS predations with measured size dependence of
  (a) $E_G$-expansion measured using STS[77] and optical method, Data –1 ($E_G = E_{PA} - W$),[78] Data –2 ($E_G = (E_{PL} + E_{PA})/2$) [79]
  (b) Core level shift of Au caped with Thiol[80] and deposited on Octan[81] shows three-dimensional features while core level shift of Au deposited on $TiO_2$[82] and Pt[83] show one-dimensional pattern.

(c) Raman acoustic frequency shift of $TiO_2$-a and $TiO_2$-b[84] $SnO_2$-a[85] nanostructures due to interparticle interaction.

(d) Raman optical frequency shift of $CeO_2$,[86] $SnO_2$-1,[87] $SnO_2$-2,[85] InP,[88] and

(e) Dielectric suppression of nanosolid silicon with Data 1, 2, and 3;[89] Data 4 and 5;[90] and Data–6.[91]

(f) Temperature and size dependence of magnetization.

## 3 Effect of oxidation

3.1 Bond-band-barrier (BBB) correlation

The BBB correlation mechanism indicates that it is necessary for an atom of oxygen, nitrogen, and carbon to hybridize its *sp* orbitals upon interacting with atoms in solid phase. Because of tetrahedron formation, non-bonding lone pairs, anti-bonding dipoles, and hydrogen-like bonds are produced, which add corresponding features to the DOS of the valence band of the host, as illustrated in Figure 7.[92] Bond forming alters the sizes and valences of the involved atoms and causes a collective dislocation of these atoms. Alteration of atomic valences roughens the surface, giving rise to corrugations of surface morphology. Charge transportation not only alters the nature of the chemical bond but also produces holes below the $E_F$ and thus creates or enlarges the $E_G$.[93] In reality, the lone-pair-induced metal dipoles often direct into the open end of a surface due to the strong repulsive forces among the lone pairs and among the dipoles as well. This dipole orientation leads to the surface dipole layer with lowered $\Phi$. For a nitride tetrahedron, the single lone pair may direct into the bulk center, which produces an ionic layer at the surface. The ionic surface network deepens the well depth, or increases the $\Phi$, as the host surface atoms donate their electrons to the electron acceptors. For carbide, no lone pair is produced but the weak antibonding feature exists due to the ion-induced polarization. However, hydrogen adsorption neither adds DOS features to the valence band nor expands the $E_G$ as hydrogen adsorption terminates the dangling bond at a surface, which minimizes the midgap impurity DOS of silicon, for instance.[34]

Figure 7 Oxygen induced *DOS* differences between a compound and the parent metal (upper) or the parent semiconductor (lower). The lone-pair polarized anti-bonding state lowers the $\Phi$ and the formation of bonding and anti-bonding generate holes close to $E_F$ of a metal or near the valence band edge of a semiconductor. For carbide, no lone pair features appear but the ion induced antibonding states will remain.

## 3.2 Experimental evidence

### 3.2.1 Surface potential barrier and bond geometry

The work function is expressed as: $\Phi = E_0 - E_F(\rho(E)^{2/3})$,[94] which is the energy separation between the vacuum level, $E_0$, and the Fermi energy, $E_F$. The $\Phi$ can be modulated by enlarging the charge density ($\rho(E)$) through lattice contraction or by raising the energy where the DOS is centered via dipole formation.[60] Dipole formation could lower the $\Phi$ of a metal surface by ~1.2 eV.[1] The work function varies from site to site with strong localized features. However, if a hydrogen-bond like forms at the surface, the $\Phi$ will restore to the original value or even higher because the metal dipoles donate the polarized electrons to the additional electronegative additives to form a '+/dipole' at the surface.[1]

Figure 8 shows a typical STM image and the corresponding model for the Cu(001)-(2×1)-$O^{2-}$ phase. The round bright spots of 0.8 ± 0.2 Å in height correspond to the lone pair induced Cu dipoles. In contrast, the protrusion for the clean Cu(110) surface is about 0.15 Å.[95] Single O-Cu-O strings are formed along the [010] direction associated with every other row missing because of the tetrahedron bond saturation.

> Figure 8 STM image [95] and the corresponding models [1] for the surface atomic valencies of Cu(110)-(2×1)-$O^{2-}$. The STM grey-scale is 0.85 Å, much higher than that of metallic Cu on a clean (110) surface (0.15 Å). The single '$O^{2-}$ : $Cu^{dipole}$ : $O^{2-}$' chain zigzagged by the non-bonding lone pairs and composed of the tetrahedron.

Dynamic XRD and very-low-energy electron diffraction (VLEED) optimization have led to quantification of atomic positions that are determined by the bond geometry such as the bond length and bond angles. A simple conversion between atomic position and bond parameters could give the bond geometry for O-Cu examples as shown in Table 3.

Table 3 Geometrical parameters for the $Cu_2O$ tetrahedron deduced from the XRD data of O-Cu(110) phase[96] and the $Cu_3O_2$ paired tetrahedra derived from VLEED calculation of O-Cu(001) surface.[97] Bond contraction with respect to the ideal length of 1.85 Å results from the effect of bond order loss. The ":" represents the lone pair interaction.

| Parameters | O-Cu(110) | O-Cu(001) | conclusion |
|---|---|---|---|
| Ionic bond length (Å) | 1.675 ($Cu^+$-$O^{2-}$) | 1.628 ($Cu^{2+}$-$O^{2-}$) | < 1.80 |
| Ionic bond length (Å) | 1.675 ($Cu^+$-$O^{2-}$) | 1.776 ($Cu^+$-$O^{2-}$) | < 1.80 |
| Lone pair length (Å) | 1.921 ($Cu^p$:$O^{2-}$) | 1.926 ($Cu^p$:$O^{2-}$) | ~ 1.92 |
| Ionic bond angle (°) | 102.5 | 102.0 | < 104.5 |
| Lone pair angle (°) | 140.3 | 139.4 | ~ 140.0 |

### 3.2.2 Valence density of states

Oxygen derived DOS features can be detected using scanning tunneling spectroscopy (STS) and ultra-violet photoelectron spectroscopy (UPS). The STS spectra in Figure 9 for an O-Cu(110) surface[95] revealed the lone pair and dipole features. Spectrum A was taken from the clean Cu(110) surface while B and C were taken from, respectively, the site above the bright spot (dipole) and the site between two bright spots along a 'O$^{2-}$ : Cu$^{dipole}$ : O$^{2-}$' chain at the Cu(110)-(2×1)-O$^{2-}$ surface. On the clean surface, empty DOS of 0.8 ~ 1.8 eV above $E_F$ are resolved and no extra DOS structures are found below $E_F$. The STS spectra recorded from the Cu(110)-(2×1)-O$^{2-}$ islands reveal that the original empty-DOS above $E_F$ are partially occupied by electrons upon chemisorption, which result in a slight shift of the empty DOS to higher energy. Additional DOS features are generated around -2.1 eV below the $E_F$. The sharp features around -1.4 eV have been detected with angular-resolved photoelectron spectroscopy (ARPES)[98] and with the de-excitation spectroscopy of metastable atoms.[99] The DOS for Cu - 3d electrons are between -2 and -5 eV,[100] and the O-Cu bonding derivatives are around the 2p-level of oxygen, -5.6 ~ -7.8 eV below $E_F$.[101] The DOS features for Cu-3d and O-Cu bonding are outside the energy range of the STS ($E_F \pm 2.5$ eV). The band-gap-expansion mechanism implies that it is possible to discover or invent new sources for light emission with a desired wavelength by controlling the extent of catalytic reaction. Intense blue-light emission from [b(Zr$_x$Ti$_{1-x}$)O$_3$ ceramics sample under Ar$^+$ ultraviolet (UV) irradiation could be direct evidence for this mechanism.[102]

It has been found that the crystal-geometry and the surface-morphology may vary from surface to surface and from material to material, the oxygen-derived DOS features are substantially the same in nature, as summarized in Table 4. The O-derived DOS features include oxygen-metal bonding (- 5 ~ -8 eV), nonbond lone-pair of oxygen (- 1 ~ -2 eV), holes of metal ions (~ $E_F$) and antibondng metal-dipole states (> $E_F$).

**Figure 9** (a) STS profiles of a Cu(110) surface [95] with and without chemisorbed oxygen. Spectra in panel (a) were obtained (A) at a metallic region, (B) on top of, and, (C) between protrusions of the 'O$^{2-}$ : Cu$^{dipole}$ : O$^{2-}$' chain [92] on the Cu(110)-(2×1)-O$^{-2}$ surface. (b) Oxygen derived DOS features (shaded areas) in the valence band of O-Pd(110) [103] and O-Cu(110) [104] surfaces. Although the microscopy and crystallography of these two systems are quite different the PES features are substantially the same. Slight difference in the feature positions results from the difference in electronegativity. Features around –1 ~-2 eV and –5 ~ -8 eV correspond to the nonbonding lone pairs and the O sp-hybrid bond states, respectively.

Table 4 Oxygen-derived DOS features adding to the valence band of metals (unit in eV). Holes are produced below $E_F$. All the data were probed with angular-resolved photoelectron spectroscopy unless otherwise indicated.

| Oxide surfaces | Methods | Anti-bond dipole $> E_F$ | Lone pair $<E_F$ | O-M bond $<E_F$ |
|---|---|---|---|---|
| O-Cu(001) [105,106] |  |  | -1.5 ± 0.5 | -6.5 ± 1.5 |
|  | VLEED [1] | 1.2 | -2.1 |  |
| O-Cu(110) [98, 104,107,108] |  | ~ 2.0 | -1.5 ± 0.5 | -6.5 ± 1.5 |
|  | STS [95] | 1.3 ± 0.5 | -2.1 ± 0.5 |  |
| O-Cu(poly) [100] |  |  | -1.5 | -6.5 ± 1.5 |
| O-Cu/Ag(110) [109] |  |  | -1.5 | -3.0; -6.0 |
| O-Rh(001) [110] | DFT | 1.0 | -3.1 | -5.8 |
| O-Pd(110) [111] |  |  | -2.0 ± 0.5 | -4.5 ± 1.5 |
| O-Gd(0001) [112] |  |  | -3.0 | -6.0 |
| O-Ru(0001) [113] |  |  | -1.0 ± 1.0 | -5.5 ± 1.5 |
| O-Ru(0001) [114] |  |  | -0.8 | -4.4 |
| O-Ru(0001) [115] | Ab initio | 1.5 | -4 | -5.5, -7.8 |
| O-Ru(0001) [116] |  | 1.7 | -3.0 | -5.8 |
| O-Ru(10$\bar{1}$0) [117] | DFT | 2.5 | -2~ -3.0 | - 5.0 |
| O-Co(Poly) [118] |  |  | -2.0 | -5.0 |
| O-diamond (001) [119] |  |  | -3.0 |  |
| (O, S)-Cu(001) [120] |  |  | -1.3 | -6.0 |
| (N, O, S)-Ag(111) [121] |  |  | -3.4 | -8.0 |

3.2.3    Lone pair interaction

A direct determination of the lone pair interaction is to measure the frequency of vibration in metal oxide surfaces using Raman and high resolution electron energy loss spectroscopy (EELS)[122,123,124] in the frequency range below 1000 cm$^{-1}$ or the shift energy of ~ 50 meV. Typical Raman spectra in Figure 10 show the lone pair vibration features in $Al_2O_3$ and $TiO_2$ powders.[125] The energy of the stretching vibration of O-M in EELS around 50 meV coincides with the energy of hydrogen bond detected using infrared and Raman spectroscopy from $H_2O$, protein, and DNA.[126] The energy for an ionic bond is normally around 3.0 eV and the energy for a Van der Waals bond is about 0.1 eV. The ~0.05 eV vibration energies correspond to the weak non-bonding interaction between the host dipole and the oxygen adsorbate.

Figure 10 Low-frequency Raman shifts indicate that weak bond interaction exists in Ti and Al oxides, which correspond to the non-bonding electron lone pairs generated during the *sp*-orbital hybridization of oxygen.

A nuclear inelastic scattering of synchrotron radiation measurement[127] revealed that additional vibrational DOS present at energies around 18 meV and 40-50 meV of oxide-caped nanocrystalline α-Fe (6 – 13 nm sizes) to the modes of the coarse-grained α-Fe. The 50 meV DOS corresponds apparently to the lone pair states while the 18 meV modes could be attributed to intercluster interaction that should increase with inverse of particle size (Figure 6c).

3.2.4  Bond forming kinetics

The spectral signatures of LEED, STM, PES/STS, TDS and EELS can be correlated to the chemical bond, surface morphology, and valence DOS and the bond strength, which enables the kinetics of oxide tetrahedron formation to be readily understood. It has been found general that an oxide tetrahedron forms in four discrete stages: (i) $O^{1-}$ dominates initially at very low oxygen dosage; (ii) $O^{2-}$-hybridization begins with lone pair and dipole formation upon second bond formation; (iii) interaction develops between lone pairs and dipoles, and finally, (iv) H-bond like forms at higher dosages. These processes give rise to the corresponding DOS features in the valence band and modify the surface morphology and crystallography, accordingly. Therefore, the events of *sp*-hybrid bonding, non-bonding lone pair, anti-bonding dipole and the H-like bonding are essential in the electronic process of oxidation, which should determine the performance of an oxide.

3.3 Summary

It is essential that an oxygen atom hybridizes its sp orbitals upon reacting with atoms in solid phase. In the process of oxidation, electronic holes, non-bonding lone pairs, anti-bonding dipoles and hydrogen-like bonds are involved, which add corresponding density-of-states features to the valence band of the host. Formation of the basic oxide tetrahedron, and consequently, the four discrete stages of bond forming kinetics and the oxygen-derived DOS features, are intrinsically common for all the analyzed systems though the patterns of observations may vary from situation to situation. What differs one oxide surface from another in observations are: (i) the site selectivity of the oxygen adsorbate, (ii) the order of the ionic bond formation and, (iii) the orientation of the tetrahedron at the host surfaces. The valencies of oxygen, the scale and geometrical orientation of the host lattice and the electronegativity of the host elements determine these specific differences extrinsically.

Knowing the bonding events and their consequences would help us to scientifically design and synthesis of oxide nanostructures with desired functions. Predictions of the functions and potential applications of the bonding events at a surface with chemisorbed oxygen are summarized in Table 5. Oxidation modifies directly the occupied valence DOS by charge transportation or polarization, in particular the band gap and work function. The involvement of the often-overlooked events of lone pair and dipoles may play significant roles in many aspects of the performance of an oxide. The bond contraction is not limited to an oxide surface but it happens at any site, where the atomic *CN* reduces.

**Table 5** Summary of special bonding events and potential applications of oxide nanomaterials.

| Events | Functions | Potential Applications |
|---|---|---|
| Anti-bonding (dipole) > $E_F$ | Work function-reduction ($\Delta\phi$) | Cold-cathode Field emission |
| Holes < $E_F$ | Band-gap expansion | PL blue-shift UV detection |
| Nonbonding (Lone pair) < $E_F$ | Polarization of metal electrons | High elasticity Raman and Far-IR low frequency activity |
| H- or CH- like bond | $\Delta\phi$ - Recovery | Bond network stabilization |
| Bond Order loss | BOLS correlation charge, mass and energy densification Cohesive energy Hamiltonian | Origin for the tunability of nano-solids |

**4    Conclusion**

The impact of the often-overlooked event of atomic CN reduction is indeed tremendous, which unifies the performance of a surface, a nanosolid and a solid in amorphous state consistently in terms of bond relaxation and its consequences on bond energy. The unusual behavior of a surface and a nanosolid has been consistently understood and systematically formulated as functions of atomic CN reduction and its derivatives on the atomic trapping potential, crystal binding intensity, and electron-phonon coupling. The properties include the lattice contraction (nanosolid densification and surface relaxation), mechanical strength (resistance to both elastic and plastic deformation), thermal stability (phase transition, liquid-solid transition, and evaporation), and lattice vibration (acoustic and optical phonons). They also cover photon emission and absorption (blue shift), electronic structures (core level disposition and work function modulation), magnetic modulation, dielectric suppression, and activation energies for atomic dislocation, diffusion, and chemical reaction. Structural miniaturization has formed indeed a new freedom that allows us to tune the physical properties that are initially non-variable for the bulk chunks by simply changing the shape and size to make use of the effect of atomic CN reduction.

The effect of size reduction and the effect of oxidation enhance each other in many aspects such as the charge localization, band gap expansion, and work function modulation. For instances, the enhancement of energy density enlarges the band gap intrinsically while charge transport from metal to oxygen enlarges the $E_G$ extrinsically by electron-hole production; both charge polarization in the process of reaction and densification in the process of size reduction lower the work function.

It is the practitioner's view that we are actually making use of the effect of bond order loss in size reduction and the effect of bond nature alteration in oxidation in dealing with oxide nanomaterials. Grasping with the factors controlling the process of bond making, breaking and relaxing would be more interesting and rewarding.

Fig.1

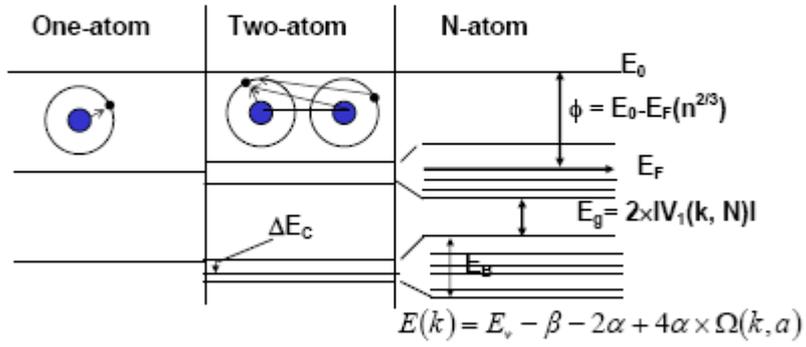

Fig-01

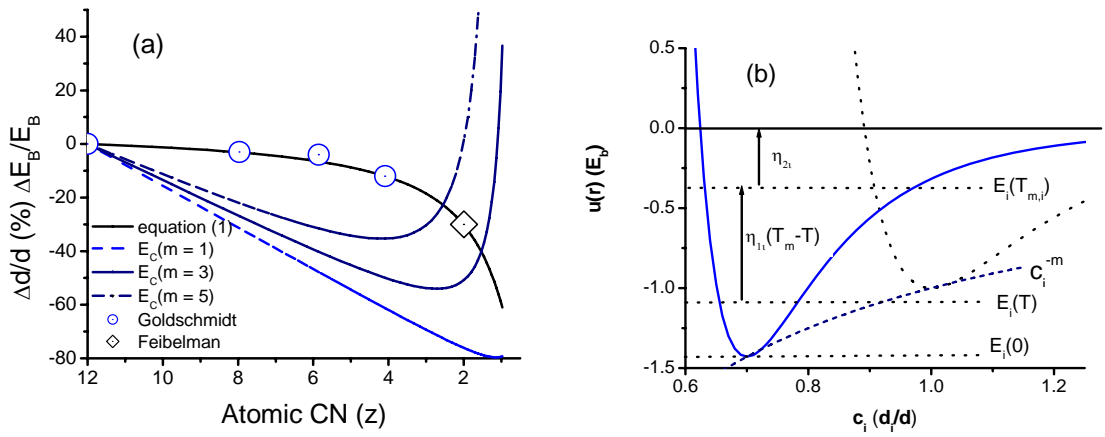

Fg-02

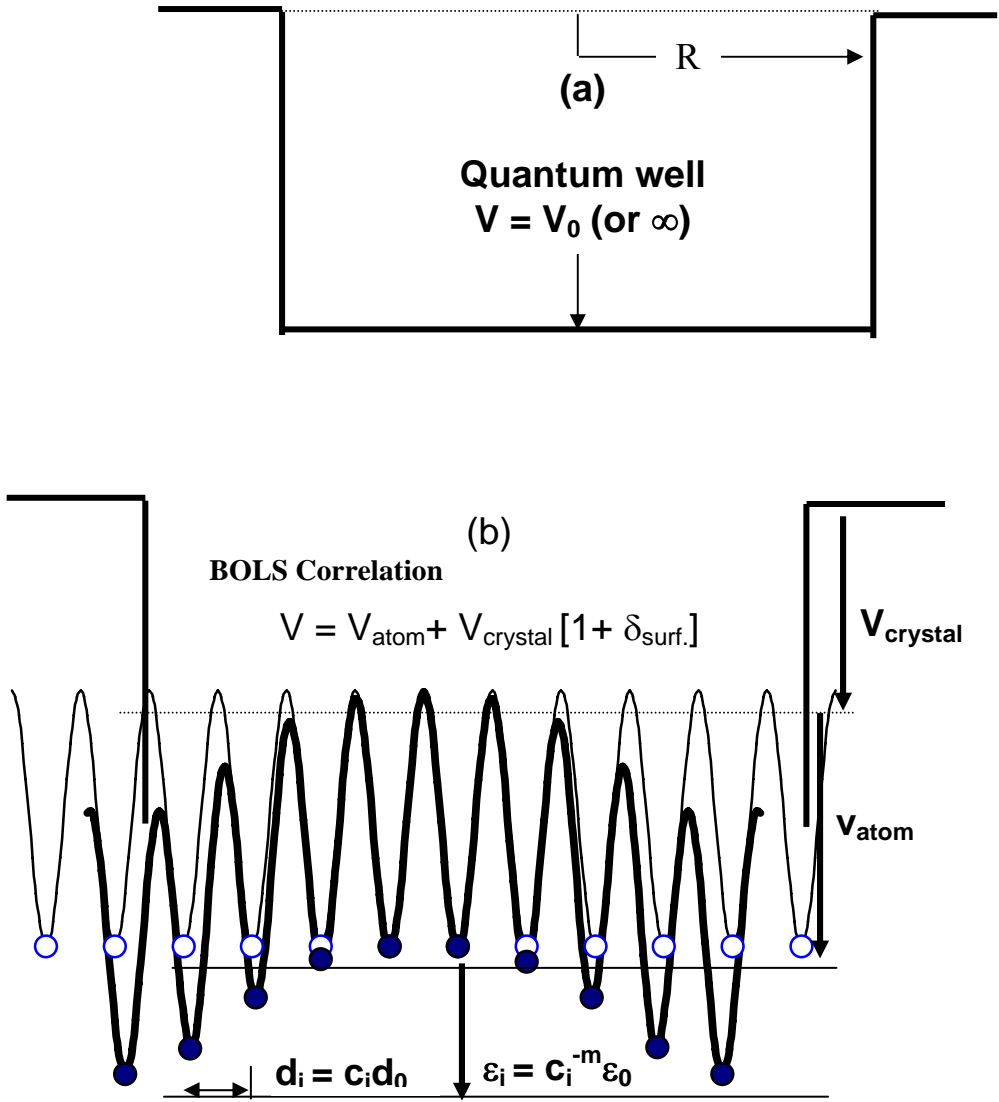

Fg-03

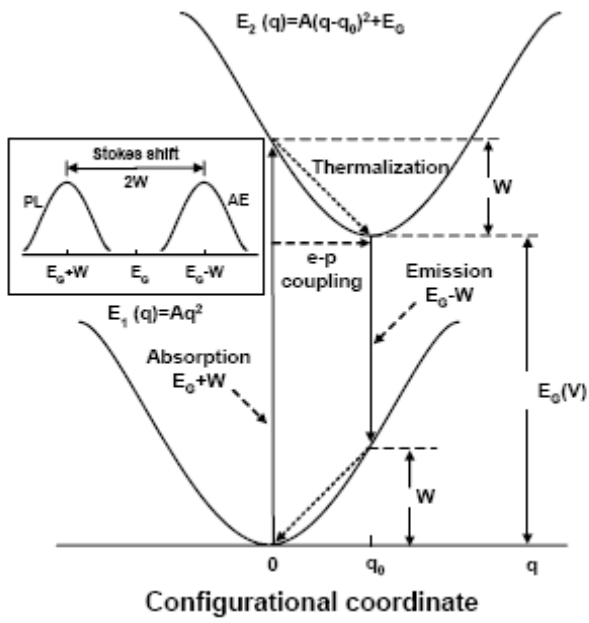

Fig-04

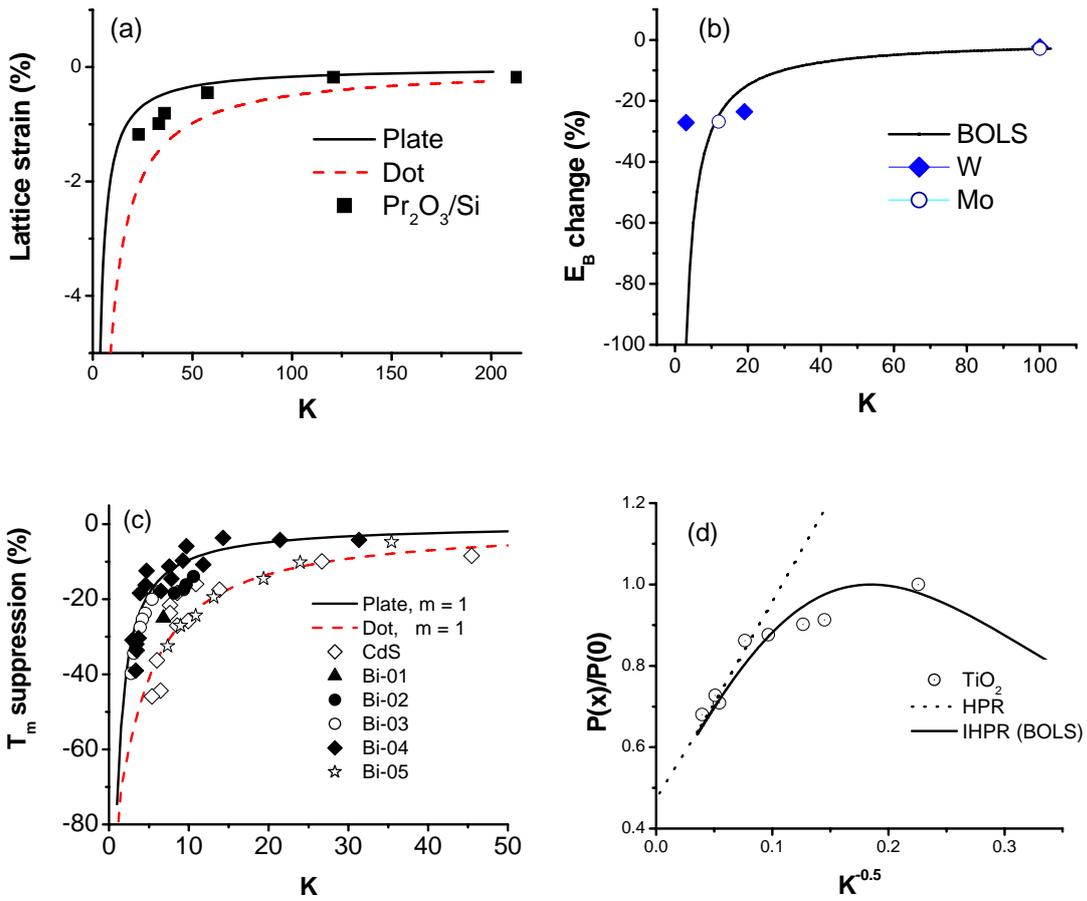

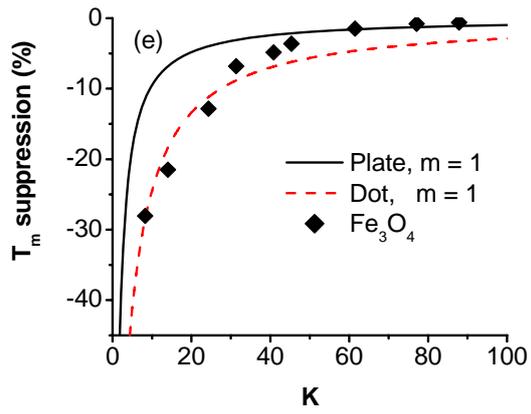
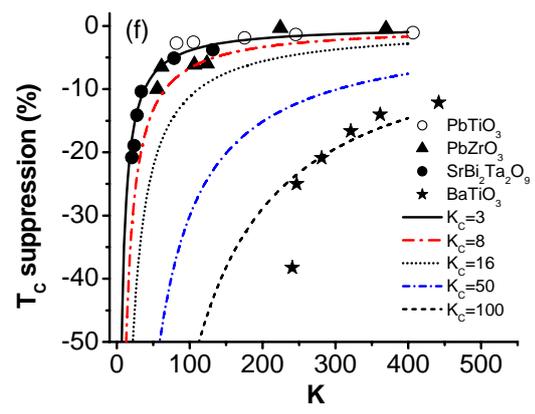

Fig-05

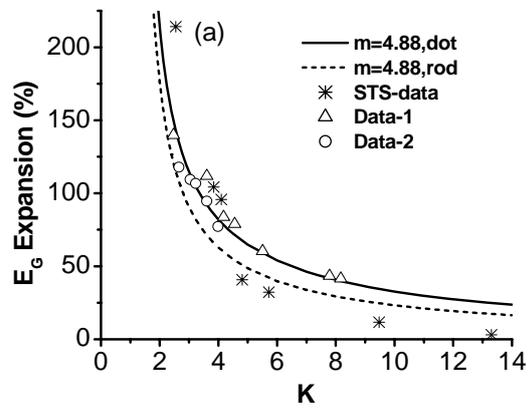
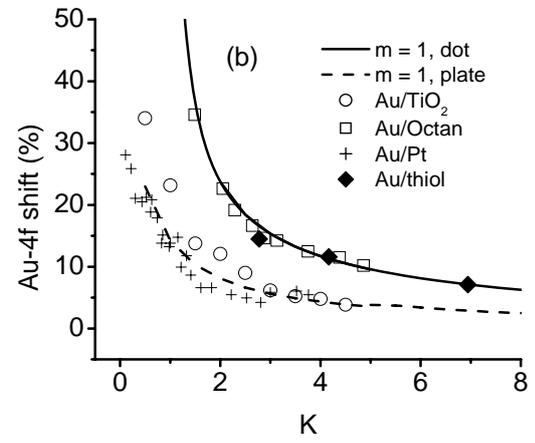

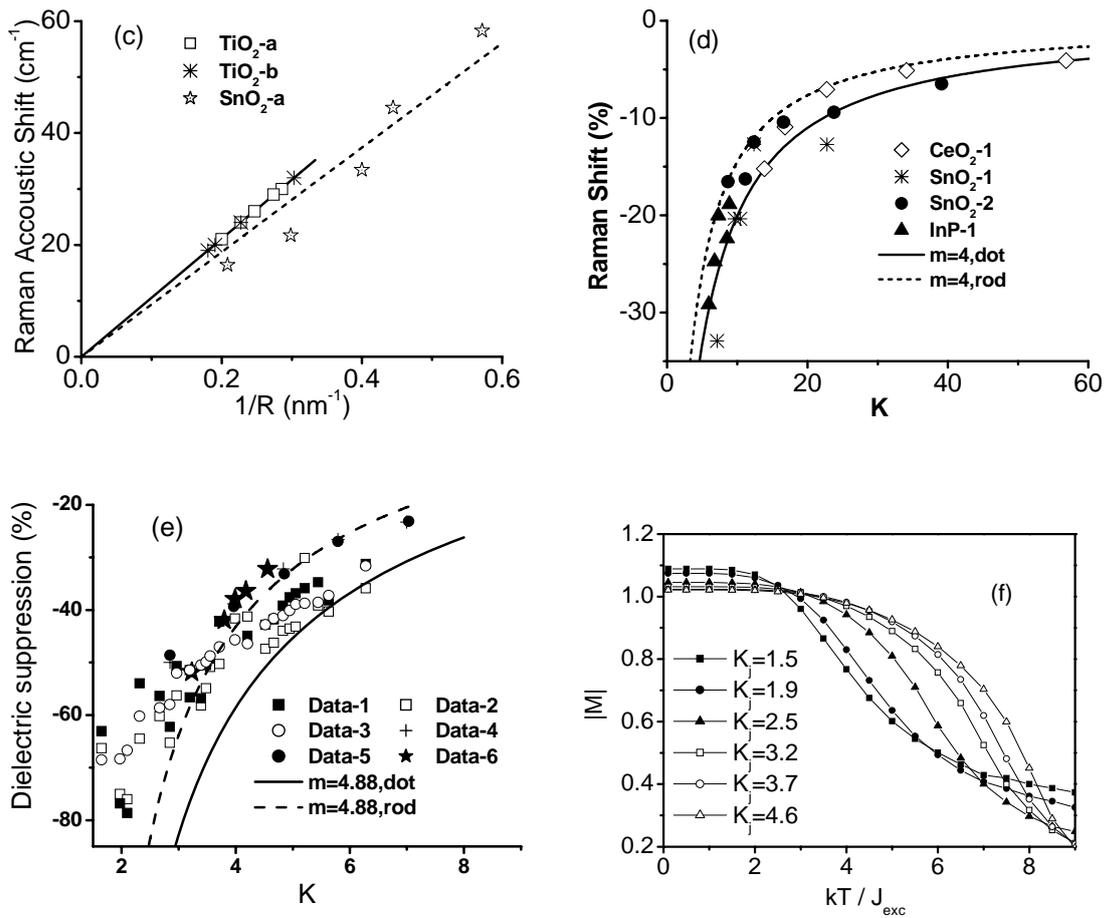

Fig-06

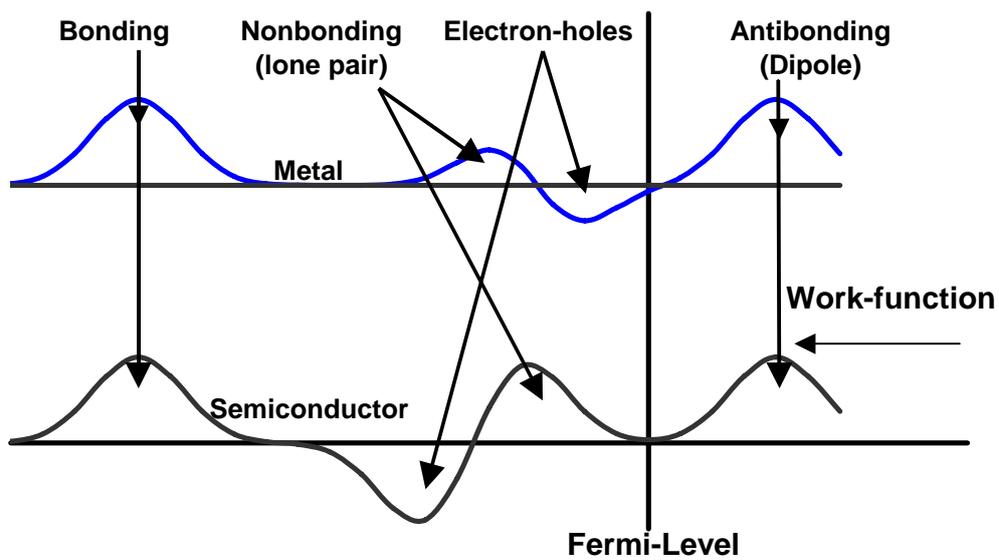

Fig-07

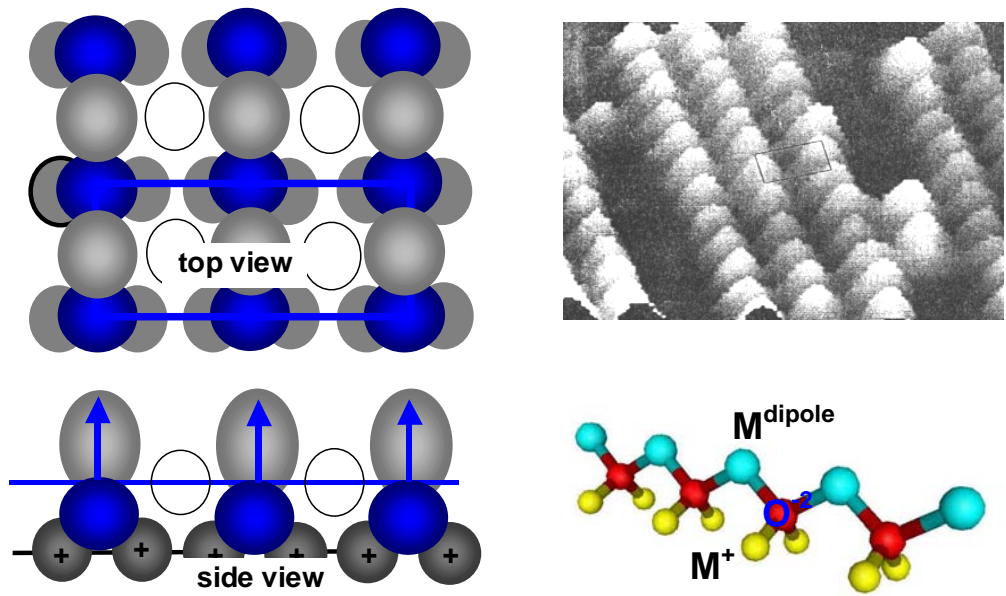

Fig-08

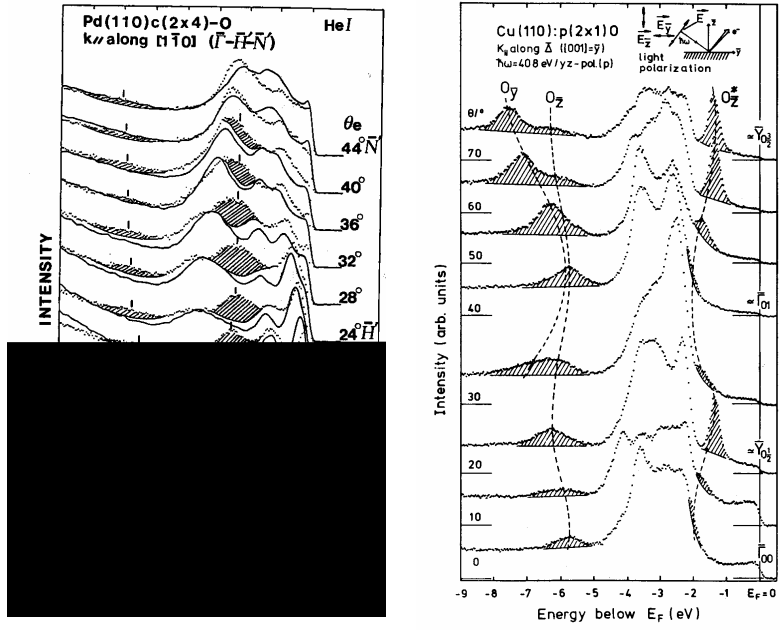

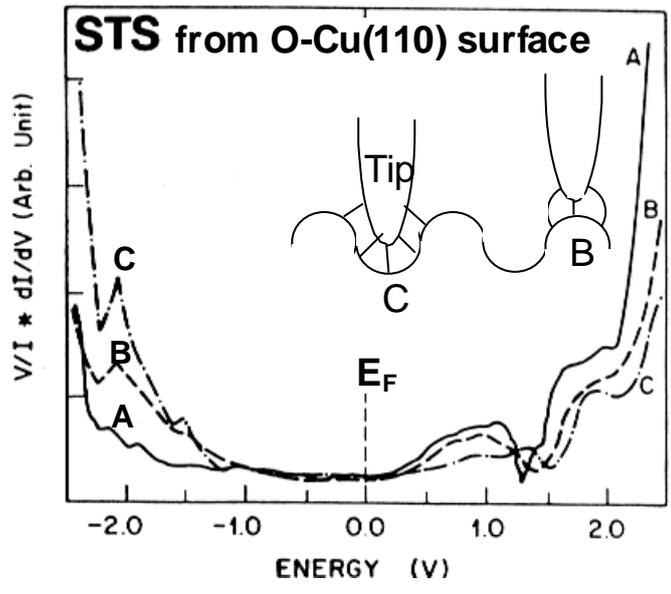

Fig-09

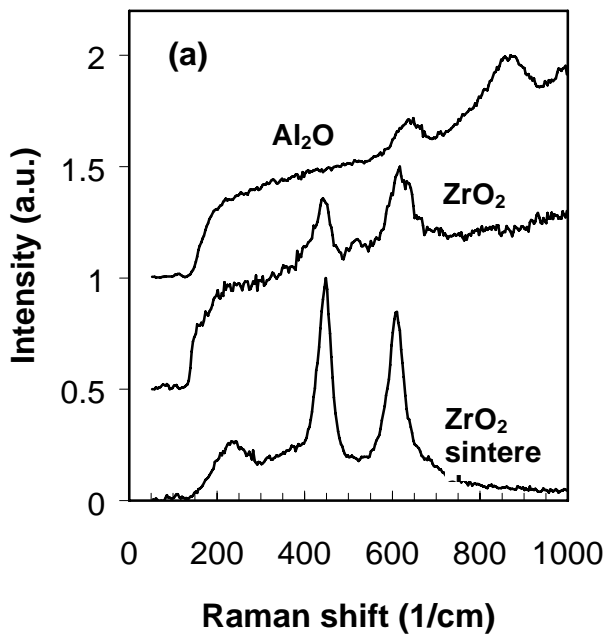

Fig.10